# On the Determination of Poisson's Ratio of Stressed Monolayer and Bilayer Submicron Thick Films

P. Martins[1], C. Malhaire[1], S. Brida[2] and D. Barbier[1]

[1] Université de Lyon, INSA-Lyon, INL, CNRS UMR 5270, Villeurbanne, F-69621, France
(Tel : +33-4-72438795; E-mail: paolo.martins@insa-lyon.fr)
[2] ESTERLINE Auxitrol SA, Esterline Sensors Group, F-18941, France
(Tel : +33-2-48667857; sbrida@auxitrol.com)

*Abstract*-In this paper, the bulge test is used to determine the mechanical properties of very thin dielectric membranes. Commonly, this experimental method permits to determine the residual stress ($\sigma_0$) and biaxial Young's modulus ($E/(1-\upsilon)$). Associating square and rectangular membranes with different length to width ratios, the Poisson's ratio ($\upsilon$) can also be determined. LPCVD $Si_3N_4$ monolayer and $Si_3N_4/SiO_2$ bilayer membranes, with thicknesses down to 100 nm, have been characterized giving results in agreement with literature for $Si_3N_4$, $E = 212 \pm 14$ GPa, $\sigma_0 = 420 \pm 8$ and $\upsilon = 0.29$.

## I. INTRODUCTION

The development of Micro Electro Mechanical Systems (MEMS) has become an economic stake since 80's and more recently the Nano Electro Mechanical Systems (NEMS) began to be developed with the downscaling trend. However, with the downscaling, accurate measurement of mechanical properties becomes a hot challenge especially since these properties may depend on the fabrication process. This may have consequences on MEMS performances and reliability [1-3]. Furthermore, architectures are more and more complex such as multilayers which make the determination of mechanical properties more difficult for each constituent material.

So far, no mechanical method exists for the simultaneous determination of the three main mechanical parameters: Young's modulus E, residual stress $\sigma_0$ and Poisson's ratio $\upsilon$, except the well-known "Bulge test" method. Indeed, the bulge test is commonly used to determine the residual stress $\sigma_0$ and the biaxial Young's modulus $E/(1-\upsilon)$ on square or circular membranes. Some authors have also shown that the association of membranes with different shapes permits the determination of the Poisson's ratio [4, 5]. However, few studies deal with the determination of the Poisson's ratio of very thin film membranes with significant results [5]. Indeed, J. S. Mitchell *et al.* [6] relate the difficulties in determining this ratio because the bulge test method is very sensitive to the geometrical errors.

In this study, an attempt was made to determine E, $\sigma_0$ and $\upsilon$ by means of bulge test on very thin ($\sim$ 100 nm) $Si_3N_4$ square and rectangular membranes with different length to width ratio (1 < b/a < 12). Few similar studies have been made on dielectric membranes with thicknesses down to 100 nm.

In this work, we have also assessed the mixture law as a general rule to extract the Young's modulus, residual stress and Poisson's ratio of each film of submicron thick $Si_3N_4/SiO_2$ bilayers.

## II. BACKGROUND

The bulge test consists in applying a pressure P on a membrane and in measuring its maximal deflection h at its center (Fig. 1).

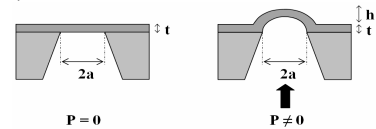

Fig. 1. Bulge test principle

Mechanical properties like Young's modulus E, the residual stress $\sigma_0$ and the Poisson's ratio $\upsilon$ can be determine from [8-10]:

$$P = C_1(a,b) \frac{t\,\sigma_0}{a^2} h + \frac{E}{12\,\alpha\,(1-\upsilon^2)} \frac{t^3}{a^4} h + f(\upsilon,a,b) \frac{t}{a^4} \frac{E}{1-\upsilon} h^3 \quad (1)$$

t represents the membrane thickness, 2a and 2b are the membrane width and length, respectively. $C_1$ (b/a) and $\alpha$ are coefficients that depend on the membrane shape and f ($\upsilon$, b/a) depends also on the membrane shape and on the Poisson's ratio. We can note that in the case of large deflections (h/t >> 1), the second term in Eq. 1 (depending on $\alpha$) can be neglected.

Lots of studies have been made to optimize the coefficient values as a function of the membrane shape and in order to take into account the particular clamping conditions of micromachined membranes [4-5] [10-13]. Tab. I presents different values of $C_1$ (b/a), f ($\upsilon$, b/a)) and $\alpha$ found in the literature as a function of the membrane shape.

In this study, the coefficients $C_1$ (b/a) and f ($\upsilon$, b/a) have been recalculated with Finite Element simulations (FE) using ANSYS software in order to verify their validity for very thin films membranes ($\sim$ 100 nm) and to compare with the literature values (Sees paragraph IV).

This experimental method is efficient to determine the residual stress $\sigma_0$ and the biaxial Young's modulus $E/(1-\upsilon)$ on thin film. Eq. 1 shows that E and $\upsilon$ are highly correlated and to find one of these parameters, the other must be assumed. To determine $\nu$ independently of E, experiments performed on square and rectangular membranes must be associated [4-6]. Indeed, it is possible to compare the ratio of the cubic coefficients in Eq. 1 to the ratio of the function f ($\nu$, b/a) for square and rectangular membranes (Eq. 2).

$$\frac{\text{Slope}_{Rect}}{\text{Slope}_{square}} = \frac{f_{Rect}(\nu, b/a)}{f_{Square}(\nu, b/a)} \left(\frac{a_{Square}}{a_{Rect}}\right)^4 \quad (2)$$

Moreover, the experimental results of J. J. Vlassak *et al.* [5], for rectangular membranes with b/a > 4, showed that membranes can be considered as infinite along the length and the deflection is independent to the aspect ratio.

The other interest of this study is to apply this method in the case of very thin multilayer films by using the simple formula of the mixture law (Eq. 3) [14-15] which can be applied for E, $\sigma_0$ and $\nu$.

$$M_{composite} = \frac{t_1}{t_{total}} M_1 + \frac{t_2}{t_{total}} M_2 + \cdots + \frac{t_n}{t_{total}} M_n \quad (3)$$

$M_{composite}$ represents either the biaxial modulus or the residual stress of the composite membrane with n layers. $t_1$, $t_2$... $t_n$ are the thicknesses of each component layer, $t_{total}$ is the multilayer thickness and $M_1$, $M_2$... $M_n$ represent either E/(1-$\nu$) or $\sigma_0$ of each layer.

### III. SAMPLE PREPARATION AND EXPERIMENTAL SETUP

#### A. Sample preparation

Dielectric membranes have been fabricated on <100> p-type, double-side polished, 100 mm silicon substrates using a standard micromachining process. Silicon nitride films have been deposited at 835 °C by LPCVD on thermally oxidized silicon substrates (Fig. 2). Two wafers issued from the same fabrication process have been processed: $Si_3N_4$ film (first wafer) with a thickness of 104 nm and bilayer $Si_3N_4/SiO_2$ film (second wafer) with a thickness of 188 nm (t($Si_3N_4$) = 90 nm and t($SiO_2$) = 98 nm). Free standing membranes have been obtained through silicon anisotropic etching in a KOH solution (Fig. 2). Several samples have been obtained with different shapes (square and rectangular). The resulting sample characteristics are summarized in Tab. II.

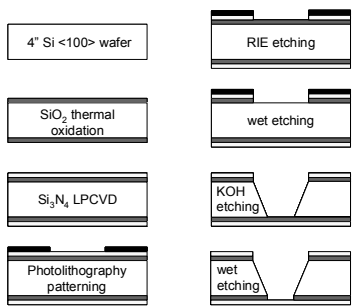

Fig. 2. Membranes fabrication: process steps

Pressure-deflection measurements (P (h)) have been performed using a WYCO NT1100 white-light interferometer microscope (Fig. 3). Pressures ranging from 10 mbar to 1 bar (depending on geometry) have been applied. Wax has been used to fix our samples on a sample holder.

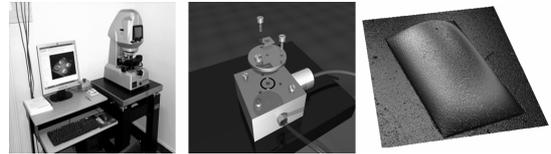

Fig. 3. Optical interferometer setup

### IV. RESULTS AND DISCUSSION

#### A. Finite Element simulations

Finite Element (FE) simulations, using ANSYS® software, have been developed in order to check if the coefficients $C_1$ (b/a) and f ($\nu$, b/a) that were found in the literature were always valid for our very thin membranes. Fig. 4 shows the evolution of these two coefficients as a function of the b/a ratio for 100 nm thick, 1 mm width membranes. An arbitrary Young's modulus value of 220 GPa and a Poisson's ratio of 0.3 have been chosen for this study.

The obtained results are in close agreement with literature values. Moreover, as shown by J. J. Vlassak *et al.* [5], for increasing b/a values from 5, $C_1$ (b/a) and f ($\nu$, b/a) become quasi independent of the b/a aspect ratio (see Fig. 4).

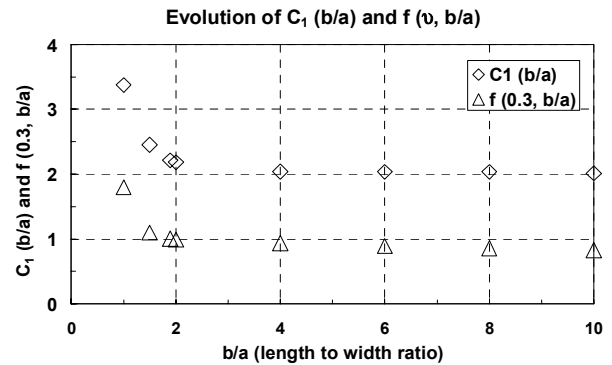

Fig. 4. Evolution of $C_1$ (b/a) and f ($\nu$, b/a) as a function of the shape b/a for 1 mm width, 100 nm thick membranes and assuming a Young's modulus of 220 GPa and a Poisson's ratio of 0.3

TABLE I
EXAMPLES OF COEFFICIENTS USED FOR DIFFERENT SHAPES

| b/a | $\alpha$ [8] | $C_1$ | f ($\nu$, b/a) | f (0.3, b/a) |
|---|---|---|---|---|
| 1 | $1.26 \times 10^{-3}$ | 3.39 [5]<br>3.45 [12]<br>3.42 [10]<br>3.39 (FE) | $(0.8+0.062\nu)^{-3}$ [5]<br>1.994(1-0.271$\nu$) [12]<br>1.91(1-0.207$\nu$) [10] | 1.82<br>1.83<br>1.79<br>1.80 (FE) |
| 2 | $2.54 \times 10^{-3}$ | 2.19 [10]<br>2.18 (FE) | 1.08(1-0.181$\nu$) [10] | 1.02 [10]<br>1.0 (FE) |
| $\infty$ | $2.6 \times 10^{-3}$ | 2 [5]<br>2 (FE) | 8/[6(1+$\nu$)] [5] | 1.02<br>0.9 (FE) |

In this study, according to the simulations, the analytical model proposed by J. J. Vlassak *et al.* is still valid for our 200 nm thick or less membranes and it was used to determine the mechanical properties of the $Si_3N_4$ monolayer and the $Si_3N_4/SiO_2$ bilayer self-standing films.

*B. Experimental results*

In the case of 2M, 3M and 4M samples, with the same geometrical parameters (Tab. II), experimental results have been obtained with a good reproducibility. Moreover, for all membranes, no hysteresis phenomenon has been observed during load and unload cycles showing a linear behavior of the membranes despite of the large induced deflections (> 90 µm).

A P/h as a function of h² normalized representation of the pressure-displacement results can be made in order to extract the y-intercept and the slope of the curves (see fig. 5 and 6). A Poisson's ratio of 0.3 for LPCVD $Si_3N_4$ was assumed according to the literature values [16] to calculate the Young's modulus E and the residual stress $\sigma_0$ for both $Si_3N_4$ and $Si_3N_4/SiO_2$ membranes. These values are summarized in Tab. II for each sample. For the monolayer $Si_3N_4$ membranes, a mean Young's modulus value of 212 ± 14 GPa and a mean residual stress of 420 ± 8 MPa have been found. For the composite $Si_3N_4/SiO_2$ bilayer membranes, the results were 147 ± 8 GPa and 107 ± 2 MPa for the Young's modulus and the residual stress, respectively.

TABLE II
RESULTS OBTAINED FOR EACH MEMBRANE USING $\nu = 0.3$

|  | n° | 2a (mm) | 2b (mm) | b/a | $\sigma_0$ (MPa) | E (GPa) |
|---|---|---|---|---|---|---|
| $Si_3N_4$ (t = 104 nm) | 1M | 3.104 | 3.104 | 1 | 439±27 | 210±16 |
|  | 2M | 2.131 | 2.131 | 1 | 400±27 | 217±19 |
|  | 3M | 2.131 | 2.131 | 1 | 409±25 | 214±16 |
|  | 4M | 2.14 | 2.14 | 1 | 429±29 | 211±18 |
|  | 5M | 1.138 | 2.131 | 1.9 | 414±34 | 219±26 |
| $Si_3N_4/SiO_2$ (t = 188 nm) | 1B | 1.89 | 1.89 | 1 | 104±8 | 150±14 |
|  | 2B | 0.662 | 0.662 | 1 | 113±9 | 153±17 |
|  | 3B | 0.750 | 0.750 | 1 | 100±8 | 156±17 |
|  | 4B | 1.39 | 7.80 | 5.6 | 103±8 | 139±15 |
|  | 5B | 0.27 | 3.28 | 12.1 | 115±10 | 145±16 |

*C. Determination of the Poisson's ratio*

Tab. III shows the different Poisson's ratio values obtained for different pairs of samples and from the analytical model proposed by Vlassak *et al.* and from Eq. 2.

TABLE III
CALCULATED POISSON RATIO FOR $Si_3N_4$ AND $Si_3N_4/SiO_2$ MEMBRANES

|  |  | $\nu$ | $\Delta\nu$ |
|---|---|---|---|
| $Si_3N_4$ | 5M/1M | 0.22 | 0.05 |
|  | 5M/2M | 0.29 | 0.07 |
|  | 5M/3M | 0.27 | 0.06 |
|  | 5M/4M | 0.24 | 0.05 |
| $Si_3N_4/SiO_2$ | 4B/1B | 0,33 | 0.05 |
|  | 4B/2B | 0.38 | 0.09 |
|  | 4B/3B | 0.41 | 0.09 |
|  | 5B/1B | 0.23 | 0.05 |
|  | 5B/2B | 0.29 | 0.08 |
|  | 5B/3B | 0.33 | 0.09 |

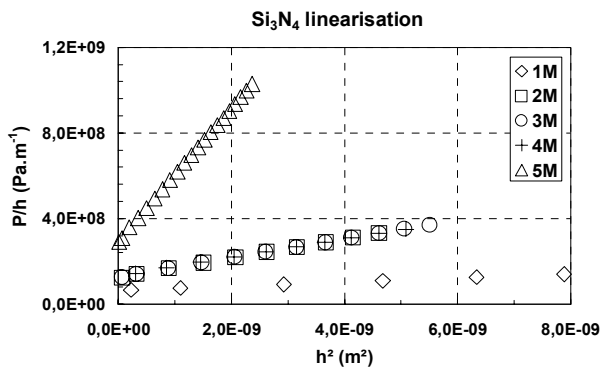

Fig. 5. Normalized pressure-displacement (P = f (h²)) curves of $Si_3N_4$ membranes

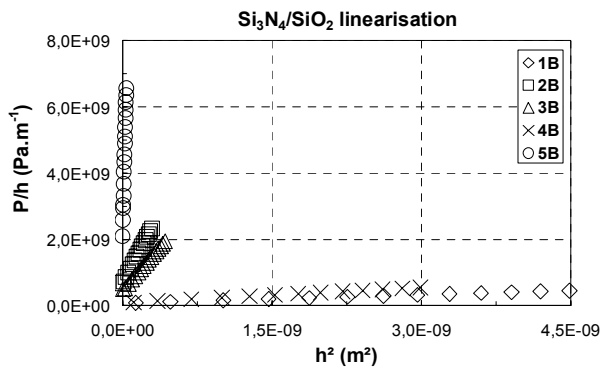

Fig. 6. Normalized pressure-displacement (P/h = f (h²)) curves of $Si_3N_4/SiO_2$ membranes

In the case of $Si_3N_4$ membranes, a Poisson's ratio scattering is observed between 0.22 and 0.29 for an expected value between 0.25 and 0.3 (LPCVD $Si_3N_4$). As regards the $Si_3N_4/SiO_2$ bilayer membranes, the composite Poisson's ratio results are more scattered and higher than for the $Si_3N_4$ monolayer (0.23 < $\nu$ < 0.41) whereas values lower than those of $Si_3N_4$ monolayer membranes were expected (except for 5B/1B samples). Even if the Poisson's ratio obtained for $Si_3N_4$ monolayers are close to the expected value, it is obvious that the determination of an accurate Poisson's ratio is very difficult, especially for bilayer membranes. Moreover, the high uncertainties in Tab. III are calculated from the lateral dimensions uncertainties showing the importance to know accurately these geometrical parameters.

Membranes that were issued from the same wafer should have the same mechanical properties but differences lower than 10 % were observed on Young's modulus results (Tab. II). These differences could come from the presence of a film thickness gradient or a stress gradient across each wafer. In this study, a mean thickness value was assumed for each wafer to calculate the mechanical properties. This may also explain the scattering on the Poisson's ratio values calculated for $Si_3N_4$. Moreover, sometimes we have observed underetching profiles on some $Si_3N_4/SiO_2$ membranes when lateral dimensions were lower than 1 mm (see Fig. 7). Even for large length/width ratio, this slightly changes the clamping conditions of the membranes and may etching influence the experimental results.

The model used to determine the Poisson's ratio is also critical. For example, for mono or multilayer membranes, the model of E. Bonnotte *et al.* leads to extreme Poisson's ratio values (>0.45) compared to that of J. J. Vlassak *et al*. This last model is the most appropriate to calculate υ.

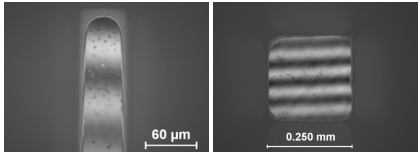

Fig. 7. $Si_3N_4/SiO_2$ Back-side membranes after anisotropic wet etching (pictures obtained using interferometer microscope)

However, when two membranes, square and rectangular, lead to very close values for the Young's modulus (we can assume that the film thickness is similar for the two samples and that the clamping conditions are good), then a precise determination of the Poisson's ratio can be made. Indeed, the 1M (square) and 5M (rectangular) $Si_3N_4$ membranes, lead to very close Young's modulus values and the calculated Poisson's ratio (Tab. II and III), in close agreement with literature values for LPCVD $Si_3N_4$ [5, 7, 16]. The same observation can be made for the 1B and 5B $Si_3N_4/SiO_2$ membranes giving a composite Poisson's ratio of 0.23.

### D. Application of mixture law

Using the mixture law (Eq. 3), an attempt was made in order to calculate the mechanical properties of thermal $SiO_2$ with our experimental results.

With E ($Si_3N_4$) ≈ 212 GPa and E ($Si_3N_4/SiO_2$) ≈ 147 GPa, we obtained E ($SiO_2$) ≈ 87 GPa. With $\sigma_0$ ($Si_3N_4$) ≈ 420 MPa and $\sigma_0$ ($Si_3N_4/SiO_2$) ≈ 107 MPa, we obtained a compressive stress $\sigma_0$ ($SiO_2$) ≈ -180 MPa. Finally, with υ ($Si_3N_4$) ≈ 0.29 and υ ($Si_3N_4/SiO_2$) ≈ 0.23, we obtained υ ($SiO_2$) ≈ 0.17.

The mechanical properties calculated on thermal $SiO_2$ are in close agreement with the literature values [17-19].

## V. CONCLUSION

These results show that the determination of E, $\sigma_0$ and υ by means of the bulge test method remains possible even for deep submicron monolayer or multilayer thin films. Large deflections can be imposed to the membranes without any plastic deformation, which simplifies the associated mechanical model. Finite Element simulations show that the coefficient values found by J. J. Vlassak *et al.* were well suited for our studied samples. But the accuracy of the results depends strongly on the geometrical parameters especially the thickness of the membranes. Young's modulus values and residual stress have been determined with accuracy better than 10 %. But the accuracy on the Poisson's ratio is about 20% in the best case. This highlights the difference between theory and experience because achieving well-controled free-standing submicron thick films is not trivial. Finally, a simple mixture law has given promising results on standard materials.